\documentclass[aip,jcp, reprint, groupedaddress, floatfix]{revtex4-2}%
\usepackage{graphicx}
\usepackage{tabularx}
\usepackage{dcolumn}
\usepackage{longtable}
\usepackage{tensor}
\usepackage{color}
\usepackage{xcolor}
\usepackage{placeins}
\usepackage{bm}
\usepackage{amsmath}
\usepackage{amsfonts}
\usepackage{amssymb}
\usepackage{enumerate}%
\providecommand{\U}[1]{\protect\rule{.1in}{.1in}}

\graphicspath{{figures/}{"C:/Users/GROUP LCPT/Desktop/LCT_SO/figures/"}{C:/Users/Jiri/Dropbox/Papers/Chemistry_papers/2021/implicit_SO_aglorithm_for_NLTDSE/figures/}}
\colorlet{RED}{red}
\begin{document}
\title{An implicit split-operator algorithm for the nonlinear time-dependent
Schr\"{o}dinger equation}
\author{Julien Roulet}
\email{julien.roulet@epfl.ch}
\author{Ji\v{r}\'{\i} Van\'{\i}\v{c}ek}
\email{jiri.vanicek@epfl.ch}
\affiliation{Laboratory of theoretical physical chemistry, Institut des sciences et
ing\'{e}nieries Chimiques, Ecole Polytechnique F\'{e}d\'{e}rale de Lausanne
(EPFL), Lausanne, Switzerland}
\date{\today}

\begin{abstract}
The explicit split-operator algorithm is often used for solving the linear and
nonlinear time-dependent Schr\"{o}dinger equations. However, when applied to
certain nonlinear time-dependent Schr\"{o}dinger equations, this algorithm
loses time reversibility and second-order accuracy, which makes it very
inefficient. Here, we propose to overcome the limitations of the explicit
split-operator algorithm by abandoning its explicit nature. We describe a
family of high-order implicit split-operator algorithms that are
norm-conserving, time-reversible, and very efficient. The geometric properties
of the integrators are proven analytically and demonstrated numerically on the
local control of a two-dimensional model of retinal. Although they are only
applicable to separable Hamiltonians, the implicit split-operator algorithms
are, in this setting, more efficient than the recently proposed integrators
based on the implicit midpoint method.

\end{abstract}
\maketitle


\section{Introduction}

The nonlinear time-dependent Schr\"{o}dinger equation (NL-TDSE) appears in the
approximate treatment of many physical processes, where the approximate
Hamiltonian depends on the state of the system. This happens, e.g., in
approximations generated by the Dirac-Frenkel variational
principle,~\cite{Dirac:1930, book_Frenkel:1934, Broeckhove_Lathouwers:1988,
book_Lubich:2008} such as the multiconfigurational time-dependent Hartree
method,~\cite{Meyer_Cederbaum:1990, Manthe_Cederbaum:1992, Beck_Jackle:2000}
variational Gaussian
approximation,\cite{Coalson_Karplus:1990,Lasser_Lubich:2020} and variational
multiconfigurational Gaussian
method,\cite{Burghardt_Giri:2008,Richings_Lasorne:2015} or in methods based on
local expansion of the potential, such as the thawed Gaussian
approximation,\cite{Heller:1975,Patoz_Vanicek:2018,Begusic_Vanicek:2020}
Hagedorn wavepacket
method,\cite{Hagedorn:1980,Faou_Lubich:2009,Lasser_Lubich:2020} or single
Hessian approximation.\cite{Begusic_Vanicek:2019,Prlj_Vanicek:2020} In
addition, many numerical methods for solving the linear Schr\"{o}dinger
equation, such as the short-iterative Lanczos
algorithm,~\cite{Lanczos:1950,Leforestier_Kosloff:1991, Park_Light:1986} can
be interpreted as exact solutions of an effective NL-TDSE.

The best known NL-TDSE is the Gross-Pitaevskii
equation,~\cite{Gross:1961,Pitaevskii:1961,Carles:2002,Carles_Sparber:2008,Minguzzi_Vignolo:2004}
which models the dynamics of Bose-Einstein
condensates.~\cite{Anderson_Cornell:1995,Dalfovo_Stringari:1999} To solve this
NL-TDSE with cubic nonlinearity, the explicit second-order split-operator
algorithm{~\cite{Feit_Steiger:1982,Kosloff_Kosloff:1983,Kosloff_Kosloff:1983a,
book_Tannor:2007}} is frequently used{~\cite{Bao_Markowich:2003}}
because it is efficient and
geometric.{~\cite{Roulet_Vanicek:2019}} However, we recently
showed~\cite{Roulet_Vanicek:2021} that the success of the explicit split-operator
algorithm in solving this NL-TDSE is due to its simple nonlinearity and,
therefore, is rather an exception than a rule. Indeed, for other
nonlinearities, the algorithm becomes time-irreversible and inefficient
because its accuracy decreases to the first order in the time
step.~\cite{Roulet_Vanicek:2021} {In many applications,
this is not an issue, but it can become a problem if accurate
wavefunctions are needed.}

As we have recently demonstrated{,}~\cite{Roulet_Vanicek:2021} an
example of a NL-TDSE, on which the
split-operator algorithm loses the second-order accuracy and time
reversibility, is provided by local control theory (LCT)%
.~\cite{Kosloff_Tannor:1989,Kosloff_Tannor:1992,Ohtsuki_Fujimura:1998,
Yamaki_Fujimura:2005,Marquetand_Engel:2006a,
Marquetand_Engel:2006b,Marquetand_Engel:2007,Engel_Tannor:2009,Bomble_Desouter-Lecomte:2011,
Vranckx_Desouter-Lecomte:2015, Vindel-Zandbergen_Sola:2016}  LCT is a
technique that aims at controlling the expectation value of a specified
operator by computing, from the state, an electric field that will either
increase or decrease the chosen expectation value. Because the electric field
is state-dependent, the interaction between this electric field and the system
is nonlinear.


To overcome the limitations of the explicit split-operator algorithm applied
to general NL-TDSEs, in our previous
work{~\cite{Roulet_Vanicek:2021}} we developed high-order
integrators by symmetrically composing the implicit midpoint method. These
integrators are applicable to the general nonlinear Schr\"{o}dinger equation
with both separable and nonseparable Hamiltonians and, in contrast to the
explicit split-operator algorithm, are efficient, while preserving the
geometric properties of the exact solution.

Here, we show that it is not necessary to abandon the split-operator algorithm
altogether, but only its explicit nature. In the linear case, the second-order
split-operator algorithms are obtained by composing two adjoint first-order
split-operator methods, which are both explicit. We show that to achieve a
second-order accuracy in the nonlinear case, one of the two adjoint algorithms
must be implicit. {Although implicit generalizations of the Verlet
algorithm
exist~\cite{Vogelaere:1956,Hairer_Wanner:2003,book_Leimkuhler_Reich:2004,
book_Hairer_Wanner:2006} for classical systems with nonseparable Hamiltonians, to the
best of our knowledge no implicit splitting methods were developed for quantum
systems with separable but nonlinear Hamiltonians. Therefore, we present an
implicit generalization of the second-order split-operator algorithm, which is
geometric, applicable to the general NL-TDSE,} and can be composed with
various composition
methods~\cite{Yoshida:1990,Suzuki:1990,Kahan_Li:1997,Sofroniou_Spaletta:2005}
to further increase its order of convergence and efficiency.

The remainder of this paper is organized as follows: In Sec.~\ref{sec:NLSE},
we present the NL-TDSE, discuss the geometric properties of its evolution
operator and describe how LCT generates a NL-TDSE. In
Sec.~\ref{sec:integrators}, we present the algorithms, their geometric
properties and the procedure employed to perform the implicit propagation
required in the implicit split-operator algorithms. In
Sec.~\ref{sec:num_examples}, we verify the convergence and the geometric
properties of the proposed integrators by performing LCT on a two-dimensional
model of retinal. Section~\ref{sec:conclusion} concludes this work.

\section{Nonlinear Schr\"{o}dinger equation}

\label{sec:NLSE}

The \emph{nonlinear time-dependent Schr\"{o}dinger equation}
\begin{equation}
i\hbar\frac{d}{dt}|\psi_{t}\rangle=\hat{H}(\psi_{t})|\psi_{t}\rangle
\label{eq:NLTDSE}%
\end{equation}
describes the time evolution of the molecular state $\psi_{t}$ under the
influence of a state-dependent Hamiltonian operator $\hat{H}(\psi_{t})$. We
will assume that the Hamiltonian
\begin{equation}
\hat{H}(\psi):=T(\hat{p})+V_{\text{tot}}(\hat{q},\psi) \label{eq:gen_nl_Ham}%
\end{equation}
is separable into a sum of a momentum-dependent kinetic energy operator
$T(\hat{p})$ and a position-dependent nonlinear potential energy operator
$V_{\text{tot}}(\hat{q},\psi)$, with $\hat{p}$ and $\hat{q}$ denoting the
momentum and position operators, respectively.

\subsection{Geometric properties of the exact evolution operator}

\label{subse:geometric_properties} The formal solution of Eq.~(\ref{eq:NLTDSE}%
) with initial condition $\psi_{t_{0}}$ can be expressed as
\begin{equation}
|\psi_{t}\rangle=\hat{U}(t,t_{0};\psi)|\psi_{t_{0}}\rangle,\qquad t\geq t_{0},
\label{eq:formal_solution}%
\end{equation}
where the exact evolution operator $\hat{U}$ is given by the time-ordered
exponential
\begin{equation}
\hat{U}(t,t_{0};\psi)=\mathcal{T}\exp\left[  -\frac{i}{\hbar}\int_{t_{0}}%
^{t}dt^{\prime}\hat{H}(\psi_{t^{\prime}})\right]  , \label{eq:U_exact}%
\end{equation}
with $\mathcal{T}$ denoting the time-ordering operator. Because the
Hamiltonian in Eq.~(\ref{eq:NLTDSE}) is nonlinear, the exact evolution
operator (\ref{eq:U_exact}) is also nonlinear; we emphasize this by including
$\psi$ as an explicit argument of $\hat{U}$. When solving Eq.~(\ref{eq:NLTDSE}%
), some geometric properties conserved by the linear time-dependent
Schr\"{o}dinger equation are not conserved in the nonlinear case. The exact
nonlinear evolution operator (\ref{eq:U_exact}) does not conserve the inner
product and, as a result, is not symplectic.~\cite{Roulet_Vanicek:2021}
Furthermore, the energy is not conserved because the state dependence of the
Hamiltonian makes it implicitly time-dependent. However, we demonstrated that
the exact nonlinear evolution operator conserves the norm and is
time-reversible.~\cite{Roulet_Vanicek:2021}

\subsection{Nonlinear Hamiltonian of local control theory}

\label{sec:LCT}

{Local control theory~\cite{Kosloff_Tannor:1989,
Kosloff_Tannor:1992}} aims at controlling the expectation value $\langle
\hat{O}\rangle_{\psi_{t}}:=\langle\psi_{t}|\hat{O}|\psi_{t}\rangle$ of a
chosen operator $\hat{O}$ in the state $\psi_{t}$. To this end, a
state-dependent electric field $\vec{E}_{\text{LCT}}(\psi_{t})$ is computed on
the fly so that the expectation value $\langle\hat{O}\rangle_{\psi_{t}}$
increases or decreases monotonously when the electric field interacts with the
system. Within the electric-dipole
approximation,~\cite{Schatz_Ratner_book:2002} this interaction is described by
the interaction potential
\begin{equation}
\hat{V}_{\text{LCT}}(\psi_{t}):=-\hat{\vec{\mu}}\cdot\vec{E}_{\text{LCT}}%
(\psi_{t}), \label{eq:nl_int_pot}%
\end{equation}
where $\hat{\vec{\mu}}$ is the electric dipole moment operator. Due to the
state-dependent control field, both the operator (\ref{eq:nl_int_pot}) and the
total potential energy operator $\hat{V}_{\text{tot}}(\psi):=\hat{V}_{0}%
+\hat{V}_{\text{LCT}}(\psi)$ [see Eq.~(\ref{eq:gen_nl_Ham})], where $\hat
{V}_{0}$ denotes the molecular potential energy operator, are nonlinear. To
control the expectation value $\langle\hat{O}\rangle_{\psi_{t}}$, the control
field used in Eq.~(\ref{eq:nl_int_pot}) is computed as
\begin{equation}
\vec{E}_{\text{LCT}}(\psi_{t}):=\pm\lambda i\langle\lbrack\hat{\vec{\mu}}%
,\hat{O}]\rangle_{\psi_{t}}^{\ast}=\mp\lambda i\langle\lbrack\hat{\vec{\mu}%
},\hat{O}]\rangle_{\psi_{t}}, \label{eq:LCT_field}%
\end{equation}
where $\lambda>0$ is a parameter that scales the amplitude of the control
field and the sign is chosen according to whether one wants to increase or
decrease $\langle\hat{O}\rangle_{\psi_{t}}$. The control field
(\ref{eq:LCT_field})
ensures~{\cite{Kosloff_Tannor:1989,Kosloff_Tannor:1992,Roulet_Vanicek:2021}%
} that the time derivative $d\langle\hat{O}\rangle_{\psi_{t}}/dt$ remains
positive [or negative, depending on the sign in Eq.~(\ref{eq:LCT_field})],
indicating a monotonic evolution of $\langle\hat{O}\rangle_{\psi_{t}}$.
However, this monotonic behavior is only guaranteed if the chosen operator
$\hat{O}$ commutes with the unperturbed molecular Hamiltonian $\hat{H}%
_{0}:=\hat{T}+\hat{V}_{0}$, i.e., if $[\hat{O},\hat{H}_{0}]=0$%
.~{\cite{Ohtsuki_Fujimura:1998,Bomble_Desouter-Lecomte:2011,Roulet_Vanicek:2021}%
}

\section{Geometric integrators for the nonlinear time-dependent
Schr\"{o}dinger equation}

\label{sec:integrators}

To solve the NL-TDSE (\ref{eq:NLTDSE}), numerical propagation methods obtain
the state $\psi_{t+\Delta t}$ at time $t+\Delta t$ from the state $\psi_{t}$
at time $t$ using the relation
\begin{equation}
|\psi_{t+\Delta t}\rangle=\hat{U}_{\text{appr}}(t+\Delta t,t;\psi)|\psi
_{t}\rangle,
\end{equation}
where $\hat{U}_{\text{appr}}(t+\Delta t,t;\psi)$ denotes an approximate
evolution operator which depends on $\psi$ and where $\Delta t$ is the
numerical time step. While all reasonable numerical methods give the exact
solution in the limit $\Delta t\rightarrow0$, some geometric properties of the
exact evolution operator may not be preserved by the numerical methods using a
finite $\Delta t$. In this section, we present the different numerical methods
and discuss their geometric properties. Detailed proofs of the geometric
properties of the presented numerical methods are shown in Appendix
\ref{sec:proof_geometric_prop}.

\subsection{Loss of geometric properties by the first-order split-operator
algorithms}

For separable Hamiltonians, the simplest split-step methods are the
\emph{explicit TV} and \emph{implicit VT} split-operator algorithms, which
approximate the exact evolution operator, respectively, as
\begin{align}
\hat{U}_{\text{TV}}(t+\Delta t,t;\psi_{t}):  &  =\hat{U}_{\hat{T}}(\Delta
t)\hat{U}_{\hat{V}_{\text{tot}}(\psi_{t})}(\Delta t) {,}%
\label{eq:U_TV}\\
\hat{U}_{\text{VT}}(t+\Delta t,t;\psi_{t+\Delta t}):  &  =\hat{U}_{\hat
{V}_{\text{tot}}(\psi_{t+\Delta t})}(\Delta t)\hat{U}_{\hat{T}}(\Delta t),
\label{eq:U_VT}%
\end{align}
where $\hat{U}_{\hat{A}}(\Delta t):=e^{-i\hat{A}\Delta t/\hbar}$ denotes an
evolution operator associated with a time-independent Hermitian operator
$\hat{A}$ and time step $\Delta t$. Both integrators (\ref{eq:U_TV}) and
(\ref{eq:U_VT}) are norm-conserving. However, both lose the symmetry and time
reversibility of the exact evolution operator. Moreover, both integrators are
only first-order accurate in the time step, and therefore, very inefficient.
Note also that the TV split-operator algorithm is explicit because it depends
on the state $\psi_{t}$ while the VT split-operator algorithm is, due to its
dependence on the state $\psi_{t+\Delta t}$, implicit and{,}
therefore{,} requires solving a nonlinear system of equations.

\subsection{Recovery of geometric properties and increasing accuracy by
composition}

Because the integrators (\ref{eq:U_TV}) and (\ref{eq:U_VT}) are adjoints of
each other (see Appendix~\ref{app:symmetry} for more details), they can be
composed together to obtain the \emph{implicit TVT} algorithm
\begin{multline}
\hat{U}_{\text{TVT}}(t+\Delta t,t;\psi_{t+\Delta t/2})\label{eq:U_TVT}\\
:=\hat{U}_{\text{TV}}(t+\Delta t,t+\Delta t/2;\psi_{t+\Delta t/2})\\
\times\hat{U}_{\text{VT}}(t+\Delta t/2,t;\psi_{t+\Delta t/2})
\end{multline}
or the \emph{implicit VTV} algorithm
\begin{multline}
\hat{U}_{\text{VTV}}(t+\Delta t,t;\psi)\label{eq:U_VTV}\\
:=\hat{U}_{\text{VT}}(t+\Delta t,t+\Delta t/2;\psi_{t+\Delta t})\\
\times\hat{U}_{\text{TV}}(t+\Delta t/2,t;\psi_{t}),
\end{multline}
depending on the order of composition. Both of these integrators are
second-order accurate in the time step and geometric because they preserve all
the geometric properties of the exact evolution operator, i.e., they are
norm-conserving, symmetric, and time-reversible. However, since both rely on
the implicit VT split-operator algorithm, both are implicit methods.

We will compare the properties of the VTV and TVT methods with the
second-order accurate \emph{implicit midpoint} method
\begin{multline}
\hat{U}_{\text{mid}}(t+\Delta t,t;\psi_{t+\Delta t/2})\label{eq:U_mid}\\
:=\hat{U}_{\text{expl}}(t+\Delta t,t+\Delta t/2;\psi_{t+\Delta t/2})\\
\times\hat{U}_{\text{impl}}(t+\Delta t/2,t;\psi_{t+\Delta t/2}),
\end{multline}
which is also geometric, and can, in contrast to the implicit TVT and VTV
split-operator algorithms, be used for both separable and nonseparable
Hamiltonians. The implicit midpoint method is obtained by composing the
first-order accurate \emph{explicit} $\hat{U}_{\text{expl}}(t+\Delta
t,t;\psi_{t}):=1-i\hat{H}(\psi_{t})\Delta t/\hbar$ and \emph{implicit}
$\hat{U}_{\text{impl}}(t+\Delta t,t;\psi_{t+\Delta t}):=[1+i\hat{H}%
(\psi_{t+\Delta t})\Delta t/\hbar]^{-1}$ Euler methods,
which are adjoints of each other. For a detailed description of the implicit
midpoint and Euler methods in the context of NL-TDSEs, we refer the reader to Ref.~\onlinecite{Roulet_Vanicek:2021}.

The second-order methods (\ref{eq:U_TVT})-(\ref{eq:U_mid}) are all symmetric
and time-reversible regardless of the size of the time step. Therefore, they
can be further composed using symmetric composition
methods{~\cite{Yoshida:1990,Suzuki:1990,Kahan_Li:1997,Abe_Kawakami:2004,Sofroniou_Spaletta:2005,book_Leimkuhler_Reich:2004,book_Hairer_Wanner:2006,Choi_Vanicek:2019,Roulet_Vanicek:2019,Roulet_Vanicek:2021}%
} in order to obtain integrators of arbitrary even orders of convergence. To
this end, starting from an integrator $\hat{U}_{p}$ of even order $p$, an
integrator $\hat{U}_{p+2}$ of order $p+2$ is generated using the symmetric
composition
\begin{multline}
\hat{U}_{p+2}(t+\Delta t,t;\psi):=\hat{U}_{p}(t+\xi_{M}\Delta t,t+\xi
_{M-1}\Delta t;\psi)\nonumber\\
\quad\cdots\hat{U}_{p}(t+\xi_{1}\Delta t,t;\psi),
\end{multline}
where $\xi_{n}:=\sum_{j=1}^{n}\gamma_{j}$ is the sum of the first $n$ real
composition coefficients $\gamma_{j}$ and $M$ denotes the total number of
composition steps. Composition coefficients $\gamma_{1},\dots,\gamma_{M}$
satisfy the relations $\sum_{n=1}^{M}\gamma_{n}=1$ (consistency),
$\gamma_{M+1-n}=\gamma_{n}$ (symmetry), and $\sum_{j=1}^{M}\gamma_{j}^{p+1}=0$
(order increase).~\cite{book_Hairer_Wanner:2006} In this work, we will use the
triple-jump~\cite{Yoshida:1990} ($M=3$) and Suzuki's
fractal~\cite{Suzuki:1990} ($M=5$) composition methods, which can both
generate integrators of arbitrary even orders of convergence. However, the
number of composition steps increases exponentially with the order of
convergence, increasing drastically the cost of performing a single time step.
To circumvent this, we will also use nonrecursive methods\cite{Kahan_Li:1997,
Sofroniou_Spaletta:2005} for obtaining sixth- eight- and tenth-order
integrators. These composition methods, which will be referred to as
\textquotedblleft optimal\textquotedblright, were designed so that they
minimize {either the sum $\sum_{n=1}^{M}|\gamma_{n}|$ or the
maximum $\max_{n}|\gamma_{n}|$ of the magnitudes of the composition} steps and, therefore, are more efficient than both the triple-jump
and Suzuki's fractal. For more details on these composition methods, see
Ref.~\onlinecite{Choi_Vanicek:2019}. {Note that by
\textquotedblleft order\textquotedblright~we mean the formal order
because, as shown by Lubich~\cite{Lubich:2008} and
Thalhammer,~\cite{Thalhammer:2012} who performed rigorous convergence analysis
of splitting methods applied to the NL-TDSE, the actual order depends on the
regularity of the initial state. Because we do not perform this analysis here,
we will verify the predicted (formal) order numerically in
Sec.~\ref{sec:num_examples}.}

\subsection{Approximate application of the explicit split-operator algorithm}

Because implicit algorithms require more expensive iterative solvers, it is
tempting to ignore the implicit character of the above-described VTV and
TVT\ algorithms, and instead employ their explicit versions, which consist of
using the state $\psi$ that is available for computing the evolution operator
for $\hat{V}_{\text{tot}}$. For example, instead of using the state
$\psi_{t+\Delta t}$ in Eq.~(\ref{eq:U_VT}) (i.e., performing the implicit
propagation exactly for the VT algorithm), the state $\psi_{\hat{T},\Delta
t/2}:=\hat{U}_{\hat{T}}(\Delta t/2)\psi_{t}$ obtained after the kinetic
propagation is often used. After composition with the TV algorithm, this
yields the \emph{approximate explicit TVT} algorithm
\begin{multline}
\hat{U}_{\text{expl TVT}}(t+\Delta t,t;{\psi_{t,\hat{T}\Delta t/2}%
})\label{eq:TVT_naive}\\
:=\hat{U}_{\text{TV}}(t+\Delta t,t+\Delta t/2;{\psi_{t,\hat{T}\Delta t/2}})\\
\times\hat{U}_{\text{VT}}(t+\Delta t/2,t;{\psi_{t,\hat{T}\Delta t/2}}).
\end{multline}
{This approximate integrator can be used for performing practical
LCT calculations in typical situations, which do not require high
accuracy.~\cite{Marquetand_Engel:2006b,
Marquetand_Engel:2007,Bomble_Desouter-Lecomte:2011,Vranckx_Desouter-Lecomte:2015,Vindel-Zandbergen_Sola:2016}%
} However, despite conserving the norm, the integrator is only first-order
accurate and neither symmetric nor time-reversible, as shown in
Ref.~\onlinecite{Roulet_Vanicek:2021}. Indeed, any explicit version of the
integrators (\ref{eq:U_TVT}) and (\ref{eq:U_VTV}) will be first-order accurate
and time-irreversible due to ignoring the implicit character of the VT algorithm.

\subsection{Solving the implicit propagation}

Both TVT and VTV implicit split-operator algorithms rely on the implicit VT
method. Using the evolution operator $\hat{U}_{VT}$ from Eq.~(\ref{eq:U_VT}),
the implicit VT propagation of a state $|\psi_{t}\rangle$ is translated into
solving the nonlinear system
\begin{equation}
\hat{U}_{\text{VT}}(t+\Delta t,t;\psi_{t+\Delta t})^{-1}| \psi_{t+\Delta
t}\rangle= |\psi_{t}\rangle.
\end{equation}
This nonlinear system can be written as $f(\psi_{t+\Delta t})=0$ with the
nonlinear functional
\begin{align}
f(\psi) :  &  = \hat{U}_{\hat{T}}(\Delta t)[\hat{U}_{\text{VT}}(t+\Delta
t,t;\psi)^{-1} \psi- \psi_{t}]\nonumber\\
&  = \hat{U}_{\hat{T}}(\Delta t)[\hat{U}_{\hat{T}}(\Delta t)^{-1}\hat{U}%
_{\hat{V}_{\text{tot}}(\psi)}(\Delta t)^{-1} \psi- \psi_{t}]\nonumber\\
&  = \hat{U}_{\hat{V}_{\text{tot}}(\psi)}(\Delta t)^{-1} \psi- \hat{U}%
_{\hat{T}}(\Delta t)\psi_{t}, \label{eq:nl_function}%
\end{align}
where we have, for convenience, included a nonzero factor of $\hat{U}_{\hat
{T}}(\Delta t)$ into the definition of $f(\psi)$.

Following Ref.~\onlinecite{Roulet_Vanicek:2021}, we employed the
Newton-Raphson method to solve this nonlinear system.
{This} method computes,
until convergence, iterative solutions of the nonlinear system using the
iterative map
\begin{equation}
\psi^{(k+1)}=\psi^{(k)}+\delta\psi^{(k)} \label{eq:iteration}%
\end{equation}
where $\psi^{(k)}$ denotes the solution obtained at the $k$th iteration and
$\delta\psi^{(k)}$ is the state obtained by solving the linear system
\begin{equation}
\hat{J}(\psi^{(k)})\delta\psi^{(k)}=-f(\psi^{(k)}), \label{eq:linear_system}%
\end{equation}
with $\hat{J}:=\delta f(\psi)/\delta\psi$ denoting the Jacobian of the
nonlinear mapping $f(\psi)$. The linear system (\ref{eq:linear_system}) is
solved using the generalized minimal residual method,~\cite{Saad_Schultz:1986,
book_Press_Flannery:1992, book_Saad:2003} an iterative method based on the
Arnoldi process\cite{Arnoldi:1951,Saad:1980} (see the supplementary material
of Ref.~\onlinecite{Roulet_Vanicek:2021} for a detailed presentation of this
algorithm). Similarly to Ref.~\onlinecite{Roulet_Vanicek:2021}, we employ the
solution from the explicit propagation, i.e., the solution obtained using
Eq.~(\ref{eq:U_TV}), as the initial guess $\psi^{(0)}$. However, if the
initial guess is too far from the implicit solution, which happens at large
time steps, the algorithm fails to converge.

The procedure described above differs from that presented in
Ref.~\onlinecite{Roulet_Vanicek:2021} only by the nonlinear system one needs
to solve. Fortunately, the use of approximations for estimating the Jacobian
is, as in Ref.~\onlinecite{Roulet_Vanicek:2021}, avoided because the Jacobian
$\hat{J}(\psi)$ of the nonlinear function (\ref{eq:nl_function}) can be
obtained analytically:
\begin{align}
\hat{J}(\psi)  &  = \frac{\delta}{\delta\psi}\left[  \hat{U}_{\hat
{V}_{\text{tot}}(\psi)}(\Delta t)^{-1} \psi\right] \nonumber\\
&  = \frac{\delta}{\delta\psi}\left[  \hat{U}_{\hat{V}_{\text{tot}}(\psi
)}(\Delta t)^{-1}\right]  \psi+ \hat{U}_{\hat{V}_{\text{tot}}(\psi)}(\Delta
t)^{-1}\hat{1}\nonumber\\
&  = \frac{i}{\hbar} \Delta t \hat{U}_{\hat{V}_{\text{tot}}(\psi)}(\Delta
t)^{-1} \frac{\delta}{\delta\psi}\left[  \hat{V}_{\text{LCT}}(\psi)\right]
\psi\nonumber\\
&  \qquad+ \hat{U}_{\hat{V}_{\text{tot}}(\psi)}(\Delta t)^{-1}\hat
{1}\nonumber\\
&  = \hat{U}_{\hat{V}_{\text{tot}}(\psi)}(\Delta t)^{-1} \left[  \hat{1} +
\frac{i}{\hbar} \Delta t \hat{V}_{\text{LCT}}(\psi)\right]  ,
\end{align}
where we employed, in the third line, the generalized complex
derivative\cite{Petersen_Pedersen:2012} of the nonlinear potential, which is
given by the bra vector
\begin{equation}
\frac{\delta}{\delta\psi}\hat{V}_{\text{LCT}}(\psi)=-\hat{\vec{\mu}}\cdot
\frac{\delta}{\delta\psi}\vec{E}_{\text{LCT}}(\psi)=\mp\lambda i\hat{\vec{\mu
}}\cdot\langle\psi|[\hat{\vec{\mu}},\hat{O}].
\end{equation}

\section{Numerical examples}

\label{sec:num_examples}

\begin{figure}
[htb]\centering\includegraphics{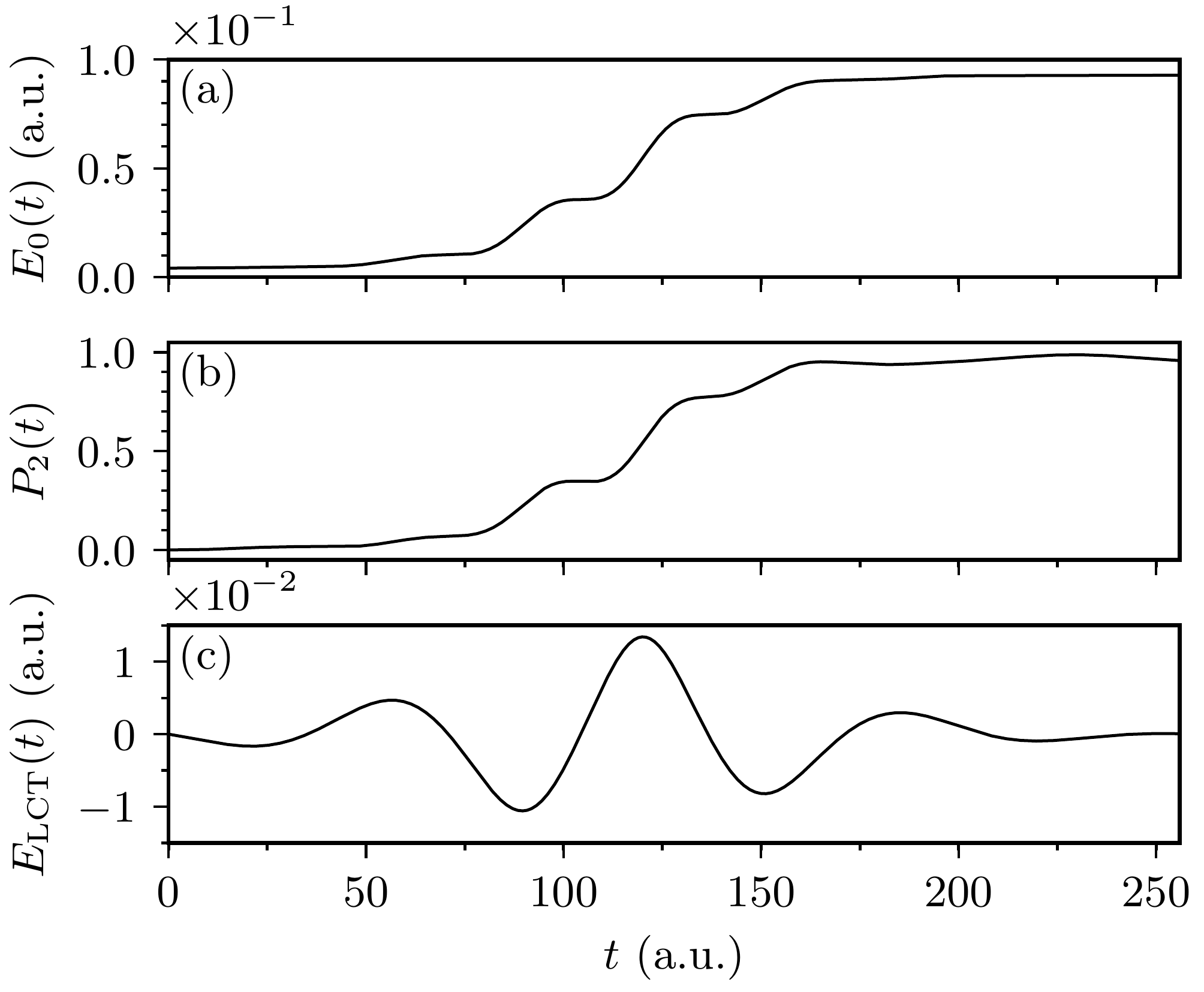}
\caption{Local control
simulation whose goal is increasing the molecular energy
$E_{0}(t)$.
{(a) Molecular energy. (b) Excited state population. (c) Pulse
obtained by LCT.}} \label{fig:pulse_pop_energy}
\end{figure}

Integrators presented in Sec.~\ref{sec:integrators} were tested on a local
control simulation in a two-dimensional model describing the
{\textit{cis-trans}} photo-isomerization of retinal. This
ultrafast reaction, which is mediated by a conical intersection, is the first
event occurring in the biological process of vision. The model, which uses the
reaction coordinate $\theta$, an angle describing the torsional motion of the
retinal molecule, and a vibronically active coupling mode $q_{c}$, was taken
from Ref.~\onlinecite{Hahn_Stock:2000} and we used it as described in
Ref.~\onlinecite{Roulet_Vanicek:2021} (see Fig.~S3 of the supplementary
material of Ref.~\onlinecite{Roulet_Vanicek:2021} for the two diabatic
potential energy surfaces of the model). We used the same grid [a regular
direct-product grid consisting of 128 points between $\theta=\pm\pi/2$~a.u.
and 64 points between $q_{c}=\pm9$~a.u.], the same initial state [a
two-dimensional Gaussian wavepacket $\psi_{0}(x)=\prod_{j=1}^{2}(\sigma
_{0,j}^{2}\pi)^{-1/4}\exp[ip_{0,j}(x_{j}-q_{0,j})/\hbar-(x_{j}-q_{0,j}%
)^{2}/2\sigma_{0,j}^{2}]$, with $x:=(\theta,q_{c})$, initial positions and
momentum $q_{0}=p_{0}=(0,0)$ a.u. and initial width $\sigma_{0}=(0.128,1)$%
~a.u., corresponding to the ground vibrational state of the harmonic fit of
the ground electronic potential energy surface], and the same initial
populations $P_{1}(0)=0.999$ and $P_{2}=0.001$ of the ground and excited
electronic states, respectively. The kinetic and potential propagations were
performed using the dynamic Fourier method\cite{Feit_Steiger:1982,
book_Tannor:2007,Kosloff_Kosloff:1983, Kosloff_Kosloff:1983a} with the Fastest
Fourier Transform in the West 3 (FFTW3) library~\cite{Frigo_Johnson:2005} to
change between position and momentum representations. Following
Ref.~\onlinecite{Roulet_Vanicek:2021}, we assumed that the electric dipole
moment operator $\hat{\vec{\bm{\mu}}}$ was coordinate independent (Condon
approximation) and aligned with the electric field $\vec{E}_{\text{LCT}}$.
Consequently, we could drop the vector symbols in Eq.~(\ref{eq:nl_int_pot}),
i.e., replace the vectors $\hat{\vec{\bm{\mu}}}$ and $\vec{E}_{\text{LCT}}$
with the scalars $\hat{\bm{\mu}}$ and $E_{\text{LCT}}$. In all simulations,
the electric dipole moment operator had unit transition (offdiagonal) elements
($\hat{\mu}_{12}=\hat{\mu}_{21}=1$ a.u.) and zero diagonal elements ($\hat
{\mu}_{11}=\hat{\mu}_{22}=0$ a.u.); which allowed the control field to couple
the two electronic states and simultaneously avoid coupling between the
vibrational states. Throughout this section, the \textbf{bold} font denotes
electronic operators expressed as $S\times S$ matrices in the basis of $S$
electronic states and that the hat $\hat{}$ denotes nuclear operators acting
on the Hilbert space of nuclear wavefunctions, i.e., square-integrable
functions of $D$ continuous degrees of freedom.

In all simulations, we used LCT for increasing the molecular energy
$E_{0}(t):=\langle\hat{\mathbf{H}}_{0}\rangle_{\psi_{t}}$ of the system, which
required employing the molecular energy operator $\hat{\mathbf{H}}_{0}$ as the
target observable. First, we performed the local control by solving the
NL-TDSE (\ref{eq:NLTDSE}) up to the final time $t_{f}=256$~a.u. using the
implicit TVT split-operator algorithm and intensity parameter $\lambda
=1.534\times10^{-1}$, which was chosen arbitrarily so that the amplitude of
the obtained control field was not too high and, at the same time, strong
enough to induce a significant increase of molecular energy (see
Fig.~\ref{fig:pulse_pop_energy}). The results indicate a successful increase
in molecular energy [panel~(a)] and, as predicted in Sec.~\ref{sec:NLSE}, this
increase is monotonic because the molecular energy operator commutes with
itself. Indeed, by construction, there are no nonzero diagonal elements in the
electric dipole moment operator and, as a result, vibrational energy cannot be
added by the control pulse ($\langle\lbrack\hat{\boldsymbol{\mu}}%
,\hat{\mathbf{T}}]\rangle_{\psi_{t}}=0$). Instead, due to the presence of
nonzero transition (offdiagonal)\ elements in the electric dipole moment, the
control pulse increases the electronic energy of our system by increasing the
excited state population $P_{2}(t)$ [panel~(b)]; this is confirmed by the
carrier frequency of the control pulse [panel~(c)], which corresponds to an
electronic transition between the two states.

\begin{figure}
[htbp]%
\centering\includegraphics{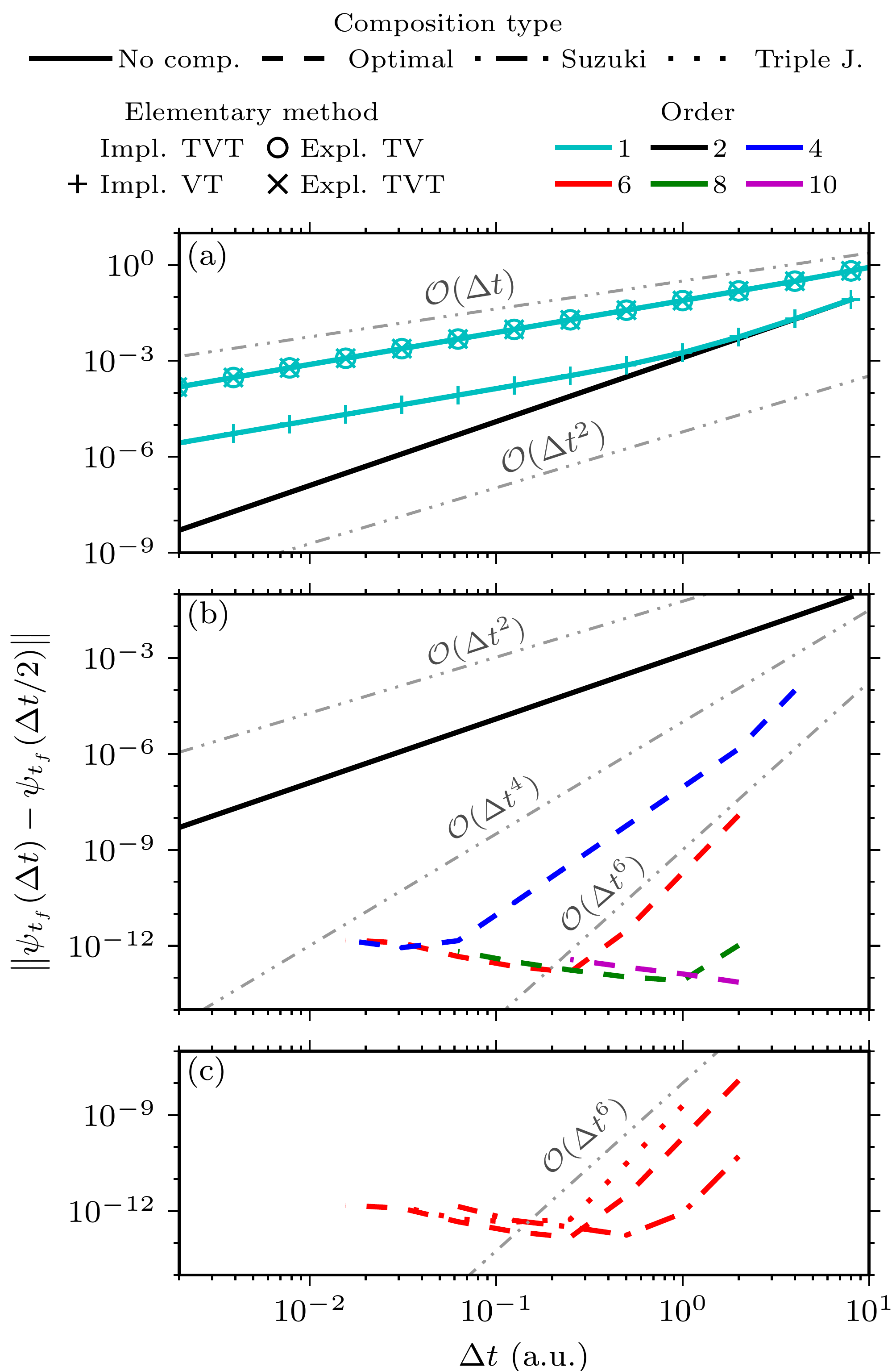}\caption{Convergence of the
molecular wavefunction at the final time $t_{f}=256$~a.u. achieved by the
local energy control. {(a) First-order and implicit TVT methods. (b) Methods obtained with
the optimal composition (Suzuki's fractal is the optimal fourth-order composition scheme~\cite{Choi_Vanicek:2019}). (c) Sixth-order methods obtained with different
composition schemes.}}\label{fig:error_vs_dt}
\end{figure}

To verify the order of convergence of the integrators presented in
Sec.~\ref{sec:integrators}, the same simulation was repeated for each
integrator with different time steps, and the errors in the obtained
wavefunctions were compared at the final time $t_{f}=256$~a.u. To measure the
convergence error, we used the $L_{2}$-norm $\Vert\psi_{t_{f}}(\Delta
t)-\psi_{t_{f}}(\Delta t/2)\Vert$ where $\psi_{t}(\Delta t)$ denotes the
wavefunction at time $t$ obtained using a time step $\Delta t$.
{Figure~\ref{fig:error_vs_dt} shows the convergence behavior of
various integrators, including higher-order integrators obtained by composing
the implicit TVT method with the triple-jump, Suzuki's fractal, and optimal
composition schemes. The results in panel (a) indicate that the implicit TVT
method has the expected order of convergence and that it is, for a given time
step, more accurate than all first-order methods, including the approximate
explicit TVT algorithm.} Comparison between different orders of the {optimal} composition of the implicit TVT algorithm [panel~(b)]
shows that, for a given time step, a higher order of composition yields more
accurate integrators. Similarly, comparing sixth-order methods obtained with
different composition schemes [panel~(c)] indicates that, for a given time
step, Suzuki's fractal composition is more accurate than both the optimal and
triple-jump compositions. {Note that after reaching a machine
precision plateau, the higher-order integrators show a slight increase in the
error with a decreasing time step, which is due to the accumulation of
roundoff errors since the number of steps increases for a fixed total time of
simulation. Moreover, some results for high-order integrators could not be
obtained because they did not converge at large time steps (the difference
between the initial guess and the implicit solution was too large for the
Newton-Raphson method to converge) and became computationally unaffordable at
smaller time steps, when the Newton-Raphson method was converging.}

\begin{figure}
[htbp]\centering\includegraphics{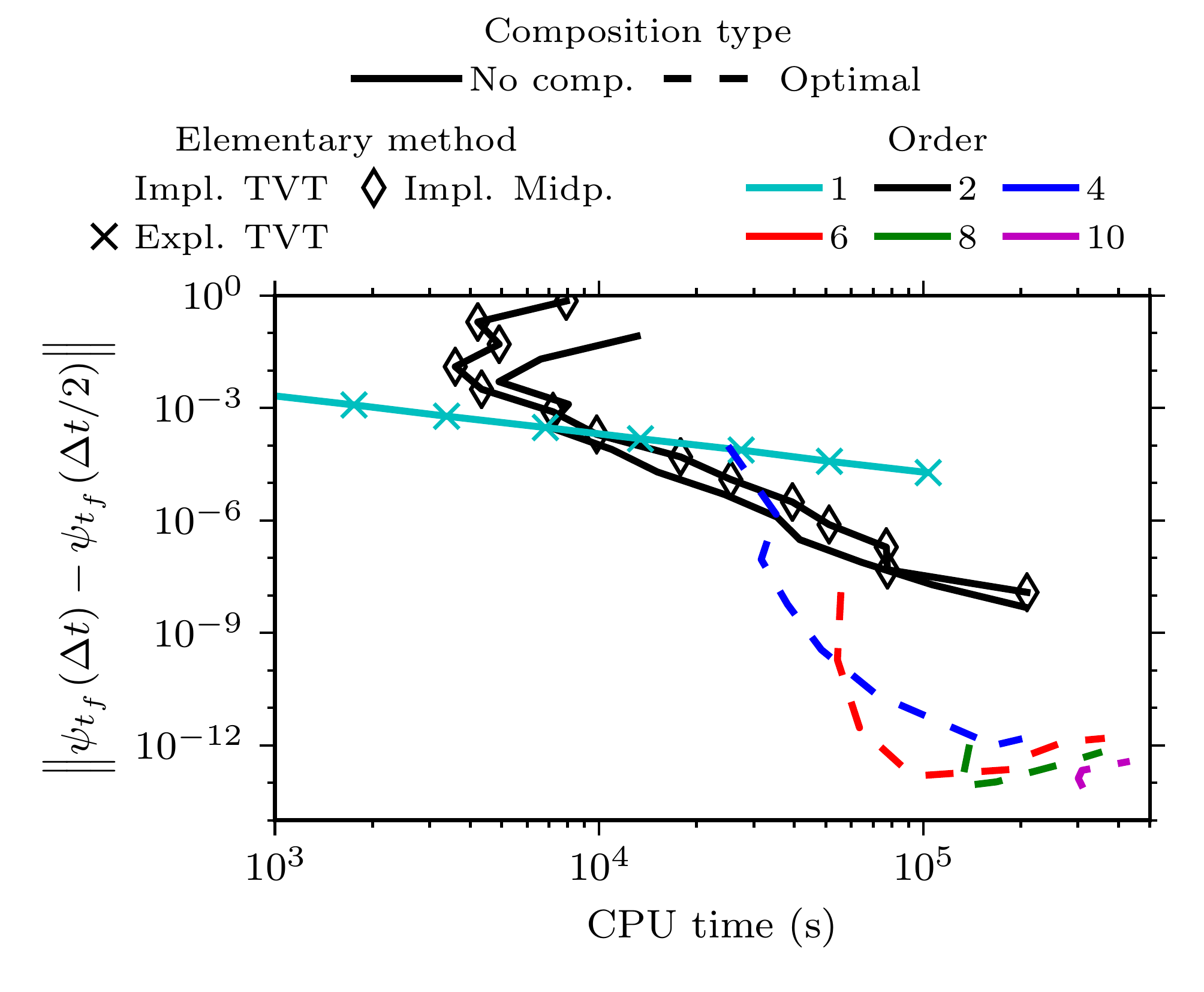}
\caption{Efficiency of {various integrators} used for simulating the local energy control of
retinal up to the final time $t_{f}=256$~a.u.} \label{fig:error_vs_CPU_time}
\end{figure}

The higher-order integrators, obtained by composition, require performing many
substeps at each time step, which increases their cost. To check that greater
accuracy for a given time step is not detrimental to efficiency, in
Fig.~\ref{fig:error_vs_CPU_time} we plot the dependence of errors of the
wavefunctions obtained by various methods on the computational cost, measured
by the central processing unit (CPU) time {(see also Fig.~S2 of the
supplementary material, which displays the efficiency results for all studied
methods)}.

{Figure~\ref{fig:error_vs_CPU_time} demonstrates} that, if high
accuracy is desired, the higher-order integrators are more efficient even
though they require performing many substeps at each time step. For example,
below an error of {$3\times10^{-4}$}, the second-order implicit
{TVT} split-operator algorithm is already more efficient than
{the approximate explicit TVT split-operator algorithm}. {Figure~\ref{fig:error_vs_CPU_time} also shows} that the implicit
TVT split-operator algorithm is more efficient than the implicit midpoint
method, indicating that the implicit split-operator algorithm is the method of
choice for separable Hamiltonians and that the implicit midpoint method should
only be used when the Hamiltonian is not separable. Indeed, for {errors
below $7\times10^{-4}$, the TVT split-operator algorithm is more efficient than} the
implicit midpoint method.

\begin{figure}
[htbp]%
\centering\includegraphics{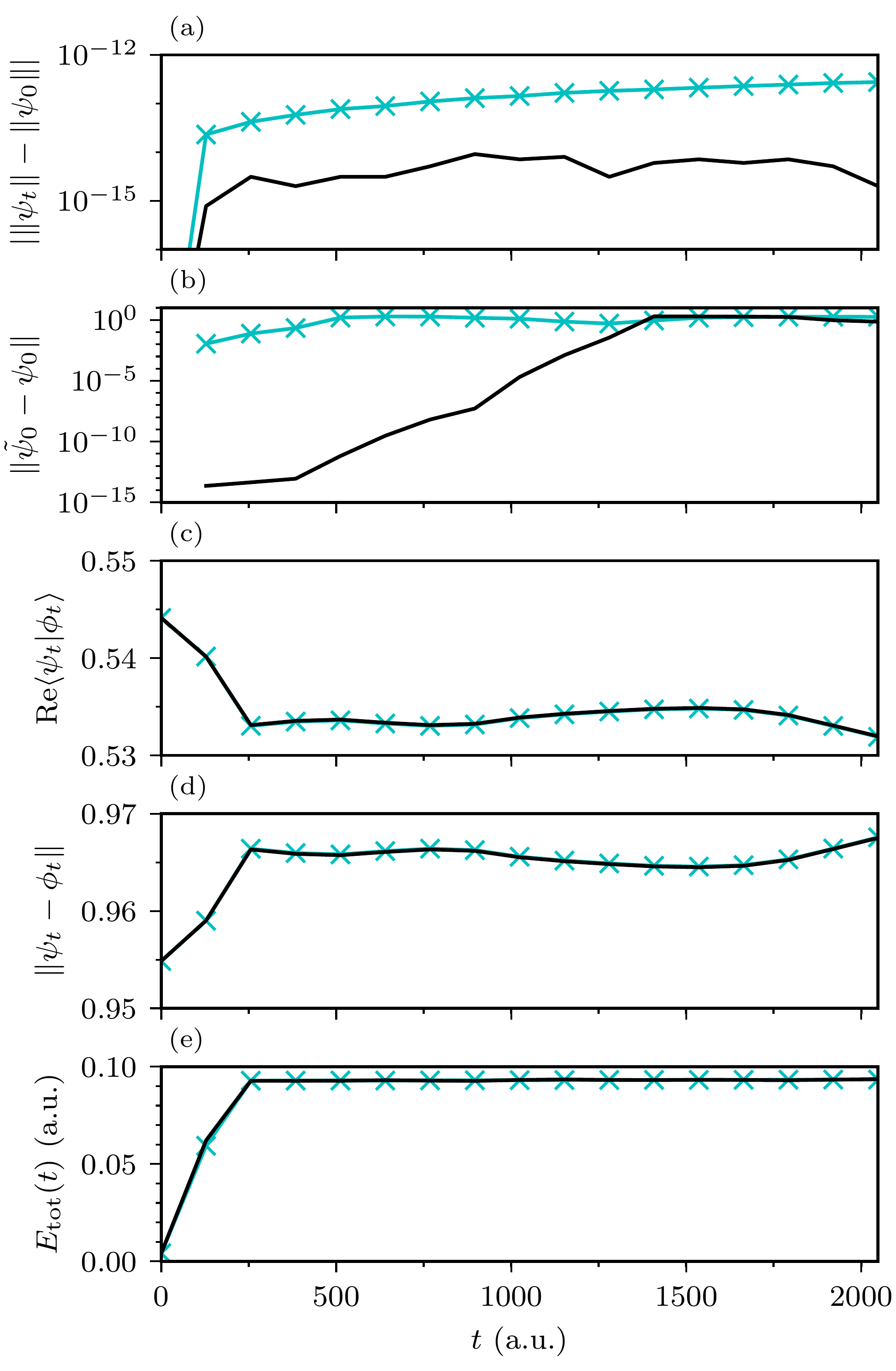}\caption{{Time dependence of the geometric properties of the {implicit and approximate explicit TVT methods} used for simulating the local energy control up to the final time $t_{f}=2048$~a.u. (a) Norm of the wavefunction. (b) Time reversibility. (c) Inner product. (d) Distance between two states ({conservation of this distance} would imply stability). (e) Total energy $E_{\text{tot}}(t):=E_{0}(t) +
\langle\hat{\mathbf{V}}_{\text{LCT}}(\psi_{t})\rangle_{\psi_{t}}$.
Time reversibility is measured by the distance between the initial state
$\psi_{0}$ and a ``forward-backward'' propagated state $\tilde{\psi}_{0}:=\hat{U}(0,t;\psi)\hat{U}(t,0;\psi)\psi_{0}$, where $\hat{U}(0,t; \psi)$ denotes a nonlinear evolution operator with a reversed time flow [see Eq.~(\ref{eq:forward_backward})].
The state $\phi_{0}$ is $\psi_{0}$ displaced along the reaction coordinate [a two-dimensional Gaussian wavepacket
with parameters $q_{0}=(0.1,0)$, $p_{0}=(0,0)$, and $\sigma_{0}= (0.128,0)$ a.u.)]. Line labels are
the same as in Fig.~\ref{fig:error_vs_dt}.}}\label{fig:geometric_properties_vs_t}%

\end{figure}

{In Fig.~\ref{fig:geometric_properties_vs_t}, we checked the
preservation of geometric properties by the implicit and approximate explicit
TVT methods (see Fig.~S3 of the supplementary material for a version of this
figure which displays the results for all the elementary methods).} Since
geometric integrators preserve geometric properties exactly regardless of the
size of the time step, we intentionally used a rather large time step $\Delta
t=2^{-2}$~a.u. We also extended the final time of the simulation to
$t_{f}=2048$~a.u. in order to induce more dynamics. Following
Ref.~\onlinecite{Roulet_Vanicek:2021}, we modified the grid to 256 points
between $\theta=\pm3\pi/2$~a.u. and 64 points between $q_{c}=\pm9$~a.u. to
ensure that the grid representation of the wavefunction at the new final time
$t_{f}$ remains converged. The results show that while {both the
implicit and approximate explicit TVT} integrators conserve the norm
[panel~(a)], only the implicit TVT {method is time-reversible
[panel~(b)].} However, due to the nonlinearity of the time-dependent
Schr\"{o}dinger equation and the accumulation of roundoff errors, one observes
a gradual loss of time reversibility as the time increases. (See Sec.~V of the
supplementary material of Ref.~\onlinecite{Roulet_Vanicek:2021} for a detailed
analysis of this loss of time-reversibility.) {The bottom three
panels of Fig.~\ref{fig:geometric_properties_vs_t} (and Fig.~S3) demonstrate
that none of the methods conserves the inner product [panel~(c)], distance
between two states [panel~(d)], or total energy [panel~(e)], because these
properties are not conserved even by the exact nonlinear evolution operator
(\ref{eq:U_exact}).}

\begin{figure}
[htbp]%
\centering\includegraphics{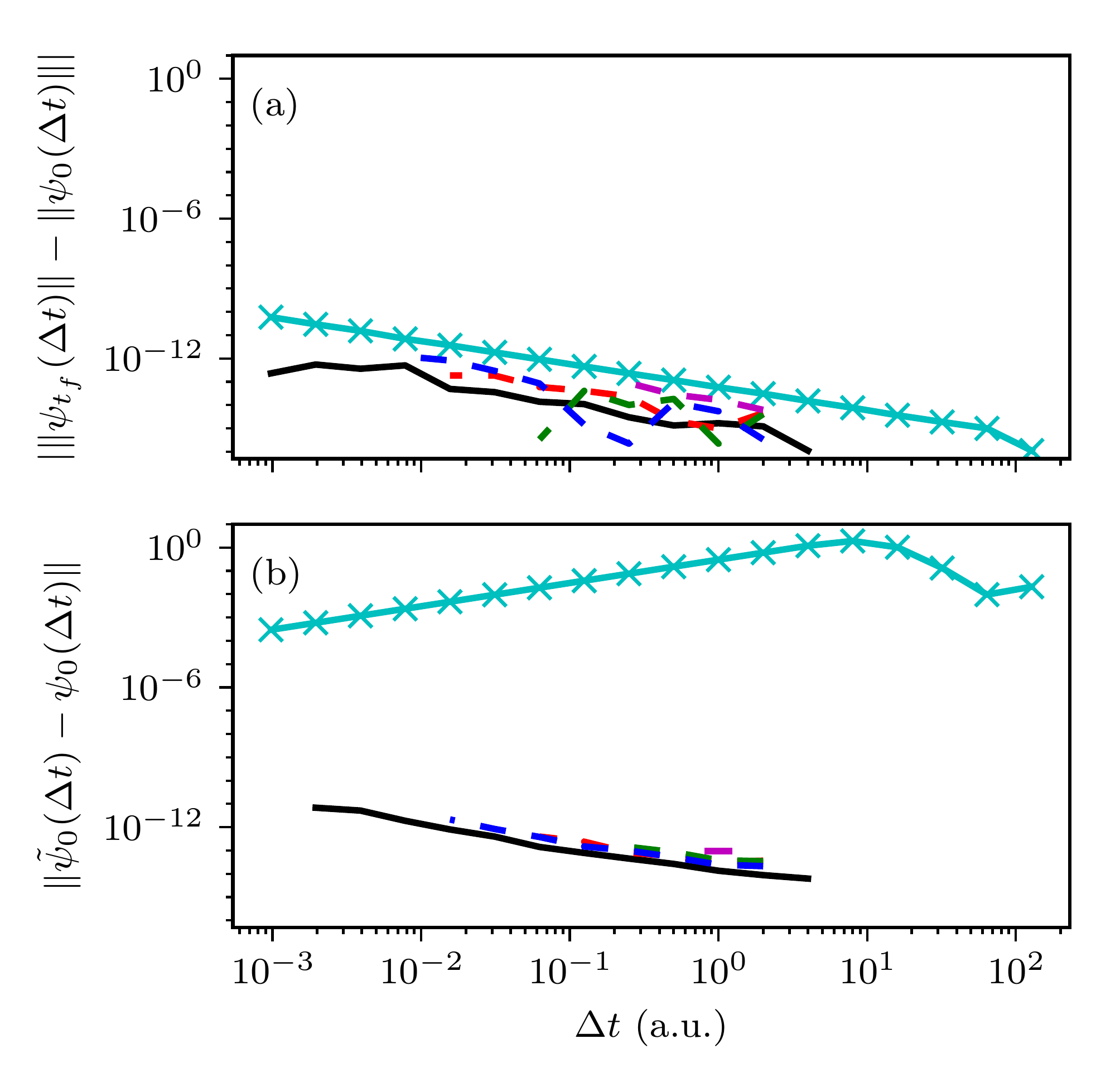}\caption{{Norm conservation (a) and time reversibility (b) of various integrators at the final time $t_{f}=256$~a.u.
as a function of the time step $\Delta t$ used for the local energy control of retinal. Reversibility is measured as in Fig.~\ref{fig:geometric_properties_vs_t} and
line labels are
the same as in Fig.~\ref{fig:error_vs_dt}.}}\label{fig:geometric_properties_vs_dt}%

\end{figure}

Figure \ref{fig:geometric_properties_vs_dt} analyze the norm conservation
[panel~(a)] and time reversibility [panel~(b)] of various integrators at the
final time $t_{f}=256$~a.u. as a function of the time step {(see
Fig.~S4 of the supplementary material for a version of this figure with all
the studied methods).} As expected, all the integrators presented in
Sec.~\ref{sec:integrators} conserve the norm, regardless of the time step.
Whereas the first-order integrators are irreversible (the time reversibility
is satisfied only to the first order in the time step), the implicit midpoint,
VTV and TVT methods as well as compositions of the latter are time-reversible
for all time steps {(see Fig.~S4)}.

\section{Conclusion}

\label{sec:conclusion}

We presented high-order integrators for solving the NL-TDSE with separable
Hamiltonians. In contrast to their first-order explicit versions, the proposed
methods, obtained by composing an implicit split-operator algorithm, preserve
all geometric properties of the exact solution: they are symmetric,
time-reversible, and norm-conserving. Moreover, the proposed integrators are
more efficient than both the explicit split-operator algorithm and the
recently proposed\cite{Roulet_Vanicek:2021} compositions of the implicit
midpoint method.

\section*{Supplementary material}

The supplementary material contains analogues of Figs.~\ref{fig:error_vs_dt}%
--\ref{fig:geometric_properties_vs_dt} of the main text that display the
numerical results for all studied methods.

\begin{acknowledgments}
The authors thank Seonghoon Choi for useful discussions and acknowledge the
financial support from the Swiss National Science Foundation within the
National Center of Competence in Research \textquotedblleft Molecular
Ultrafast Science and Technology\textquotedblright\ (MUST) and from the
European Research Council (ERC) under the European Union's Horizon 2020
research and innovation program (grant agreement No. 683069 -- MOLECULE).
\end{acknowledgments}

\section*{Author declarations}

\subsection*{Conflict of interest}

The authors have no conflicts to disclose.

\section*{Data availability}

{The data that support the findings of this study are openly
available in Zenodo at http://doi.org/10.5281/zenodo.5566833.}

\appendix

\section{Geometric properties of various integrators}

\label{sec:proof_geometric_prop}

Here we demonstrate the geometric properties of the explicit TV and implicit
VT, TVT, and VTV algorithms. We refer the reader to the Appendix of
Ref.~\onlinecite{Roulet_Vanicek:2021} for the analogous proofs for the
approximate explicit TVT algorithm and the implicit midpoint method.

\subsection{Norm conservation}

The evolution operator $\hat{U}_{\hat{T}}(\Delta t)$ of the Hermitian kinetic
energy operator $\hat{T}$ conserves the norm $\Vert\psi_{t}\Vert$ of the state
$\psi_{t}$ because
\begin{align}
\Vert\hat{U}_{\hat{T}}(\Delta t)\psi_{t}\Vert^{2}  &  =\langle\psi_{t}|\hat
{U}_{\hat{T}}(\Delta t)^{\dagger}\hat{U}_{\hat{T}}(\Delta t)\psi_{t}%
\rangle=\langle\psi_{t}|\psi_{t}\rangle\nonumber\\
&  =\Vert\psi_{t}\Vert^{2}, \label{eq:U_T_norm}%
\end{align}
where we used the relation
\begin{equation}
\hat{U}_{\hat{A}}(\Delta t)^{\dagger}=(e^{-i\hat{A}\Delta t/\hbar})^{\dagger
}=\hat{U}_{\hat{A}}(\Delta t)^{-1}, \label{eq:Herm_evol}%
\end{equation}
which holds for any Hermitian operator $\hat{A}$, to obtain the second equality.

For the potential evolution operator, we first assume that while the operator
$\hat{V}_{\text{tot}}:\psi\mapsto\hat{V}_{\text{tot}}(\psi)\psi$ is nonlinear,
for each $\phi$ the operator $\hat{V}_{\text{tot}}(\phi):\psi\mapsto\hat
{V}_{\text{tot}}(\phi)\psi$ is linear. Moreover, we assume that $\hat
{V}_{\text{tot}}(\phi)$ has real expectation values $\langle\hat
{V}_{\text{tot}}(\phi)\rangle_{\psi}$ in any state $\psi$, which for a linear
operator implies that it is Hermitian. Therefore, the evolution operator
$\hat{U}_{\hat{V}_{\text{tot}(\phi)}}(\Delta t)$ conserves, for any $\phi$,
the norm of the state $\psi_{t}$ because
\begin{align}
\Vert\hat{U}_{\hat{V}_{\text{tot}(\phi)}}(\Delta t)\psi_{t}\Vert^{2}  &
=\langle\psi_{t}|\hat{U}_{\hat{V}_{\text{tot}(\phi)}}(\Delta t)^{\dagger}%
\hat{U}_{\hat{V}_{\text{tot}(\phi)}}(\Delta t)\psi_{t}\rangle\nonumber\\
&  =\langle\psi_{t}|\psi_{t}\rangle=\Vert\psi_{t}\Vert^{2},
\end{align}
where we used Eq.~(\ref{eq:Herm_evol}) to obtain the second equality.

Composing two norm-conserving evolution operators $\hat{U}_{\hat{A}}$ and
$\hat{U}_{\hat{B}}$ of Hermitian operators $\hat{A}$ and $\hat{B}$,
respectively, yields a norm-conserving integrator $\hat{U}_{\hat{A}\hat{B}%
}(\Delta t):=\hat{U}_{\hat{A}}(\Delta t)\hat{U}_{\hat{B}}(\Delta t)$. Indeed,
we have
\begin{align}
\Vert\hat{U}_{\hat{A}\hat{B}}(\Delta t)\psi_{t}\Vert^{2}  &  =\Vert\hat
{U}_{\hat{A}}(\Delta t)\psi_{t}^{\prime}\Vert^{2}=\Vert\psi_{t}^{\prime}%
\Vert^{2}\nonumber\\
&  =\Vert\hat{U}_{\hat{B}}(\Delta t)\psi_{t}\Vert^{2}=\Vert\psi_{t}\Vert^{2},
\end{align}
where $\psi_{t}^{\prime}:=\hat{U}_{\hat{B}}(\Delta t)\psi_{t}$. Therefore, all
proposed integrators (including the integrators obtained by symmetric
composition of TVT or VTV algorithms) conserve the norm because they are all
compositions of the norm-conserving integrators $\hat{U}_{\hat{T}}(\Delta t)$
and $\hat{U}_{\hat{V}_{\text{tot}(\phi)}}(\Delta t)$.

\subsection{Symmetry and time-reversibility}

\label{app:symmetry}

{ In the theory of dynamical systems, an \emph{adjoint} $\hat{U}_{\text{appr}%
}(\psi)^{\ast}$} of $\hat{U}_{\text{appr}}(\psi)$ is defined as the inverse of
the evolution operator taken with a reversed time flow:
\begin{equation}
\hat{U}_{\text{appr}}(t+\Delta t,t;\psi)^{\ast}:=\hat{U}_{\text{appr}%
}(t,t+\Delta t;\psi)^{-1}.
\end{equation}
If the evolution operator is equal to its adjoint, i.e., if $\hat{U}%
(t,t_{0};\psi)=\hat{U}(t,t_{0};\psi)^{\ast}$, the evolution operator $\hat
{U}(t,t_{0};\psi)$ is said to be symmetric. Time reversibility results from
symmetry because for a symmetric evolution operator, propagating an initial
state $\psi_{t_{0}}$ forward to time $t$ and then backward to time $t_{0}$,
recovers $\psi_{t_{0}}$, i.e.,
\begin{align}
&  \hat{U}_{\text{appr}}(t_{0},t;\psi)\hat{U}_{\text{appr}}(t,t_{0};\psi
)\psi_{t_{0}}\nonumber\\
&  =\hat{U}_{\text{appr}}(t_{0},t;\psi)^{\ast}\hat{U}_{\text{appr}}%
(t,t_{0};\psi)\psi_{t_{0}}\nonumber\\
&  =\hat{U}_{\text{appr}}(t,t_{0};\psi)^{-1}\hat{U}_{\text{appr}}(t,t_{0}%
;\psi)\psi_{t_{0}}=\psi_{t_{0}}. \label{eq:forward_backward}%
\end{align}
Neither the explicit TV nor implicit VT method is symmetric because
\begin{align}
\hat{U}_{TV}(t+\Delta t,t;\psi_{t})^{\ast}  &  =\hat{U}_{TV}(t,t+\Delta
t;\psi_{t+\Delta t})^{-1}\nonumber\\
&  =[\hat{U}_{\hat{T}}(-\Delta t)\hat{U}_{\hat{V}_{\text{tot}}(\psi_{t+\Delta
t})}(-\Delta t)]^{-1}\nonumber\\
&  =\hat{U}_{\hat{V}_{\text{tot}}(\psi_{t+\Delta t})}(\Delta t)\hat{U}%
_{\hat{T}}(\Delta t)\nonumber\\
&  =\hat{U}_{VT}(t+\Delta t,t;\psi_{t+\Delta t})\nonumber\\
&  \neq\hat{U}_{TV}(t+\Delta t,t;\psi_{t}) \label{eq:U_TV_sym}%
\end{align}
and
\begin{align}
\hat{U}_{VT}(t+\Delta t,t;\psi_{t+\Delta t})^{\ast}  &  =\hat{U}%
_{VT}(t,t+\Delta t;\psi_{t})^{-1}\nonumber\\
&  =[\hat{U}_{\hat{V}_{\text{tot}}(\psi_{t})}(-\Delta t)\hat{U}_{\hat{T}%
}(-\Delta t)]^{-1}\nonumber\\
&  =\hat{U}_{\hat{T}}(\Delta t)\hat{U}_{\hat{V}_{\text{tot}}(\psi_{t})}(\Delta
t)\nonumber\\
&  =\hat{U}_{TV}(t+\Delta t,t;\psi_{t})\nonumber\\
&  \neq\hat{U}_{VT}(t+\Delta t,t;\psi_{t+\Delta t}). \label{eq:U_VT_sym}%
\end{align}
As a result, neither the TV nor VT method is time-reversible.

From Eqs.~(\ref{eq:U_TV_sym}) and (\ref{eq:U_VT_sym}), we notice that the
explicit TV and VT methods are, in fact, adjoints of each other, i.e.,
$\hat{U}_{TV}^{\ast}=\hat{U}_{VT}$ and $\hat{U}_{VT}^{\ast}=\hat{U}_{TV}$. In
general, the composition of adjoint methods $\hat{U}$ and $\hat{U}^{\ast}$,
each with a time step $\Delta t/2$, yields symmetric methods $\hat{U}\hat
{U}^{\ast}$ and $\hat{U}^{\ast}\hat{U}$.~\cite{book_Hairer_Wanner:2006}
Because the implicit TVT and VTV methods are such compositions of adjoint TV
and VT methods, both TVT and VTV integrators are symmetric. Applying symmetric
composition schemes to these symmetric methods will always yield a symmetric
method.~\cite{book_Hairer_Wanner:2006} Therefore, the proposed high-order
integrators are also symmetric and time-reversible.

\bibliographystyle{aipnum4-2}
\bibliography{LCT_SO}

\end{document}